\documentstyle[aps]{revtex}
\begin{document}
\title{ Centrifugal Effects in a Bose-Einstein Condensate
in the TOP-Magnetic Trap}
\author{A. B. Kuklov$^{1,2}$, N. Chencinski$^1$, A. M. Levine$^1$, 
W. M. Schreiber$^1$, and Joseph L. Birman$^2$}
\address{$^1$ Department of Applied Sciences, 
The College of  Staten Island, CUNY,
     Staten Island, NY 10314}
\address{$^2$ Department of Physics, The City College, CUNY, New York,
NY 10031}

\maketitle
   
\begin{abstract}
 Single particle states in the atomic trap employing the
rotating magnetic field  
are found using the full time-dependent instantaneous trapping potential. 
These states are compared  with those of the effective time-averaged
 potential. 
We show that the trapping is possible when the frequency of the rotations
exceeds some threshold. Slightly above this threshold the weakly interacting
gas of the trapped atoms acquires the properties of a quasi-1D system in
the frame rotating together with the field. 
The role of the atom-atom interaction in changing the ideal gas solution
is discussed. We show that in the limit of large numbers of particles
the rotating field can be utilized as a driving force
 principally for the center of mass motion as well   as for the angular momentum
$L = 2$ normal modes of the Bose condensate.
A mechanism of quantum evaporation forced by the rotating field is
analyzed.
\\
   
\end{abstract}

   \section{Introduction}
   
The novel methods \cite{ANDERSON,BRADLEY,DAVIS} for storing atoms
at very low densities and temperatures open up new opportunities
for studying the role of the atom-atom interaction in macroscopic quantum
phenomena \cite{BEC}. The problems of Bose-Einstein condensate
formation \cite{SUBIR,KAGAN,STOOF}, the dynamical response of the condensate
in the trap \cite{STRIN1}, and the interaction of the condensate with light 
\cite{SHLAP} can now be investigated experimentally.
  
  The zero dimensional geometry and the small size of the atomic traps
restrict the direct observation of the most spectacular effects known
from the history of superfluidity of  HeII (see in, e.g., \cite{REV}).
 Therefore devising new practical methods for probing the condensate  
in the atomic traps becomes of crucial importance. 
In this regard the recent suggestion
\cite{STRIN2} to analyze the rotational properties of the trapped 
atomic cloud appears to be very promising.
As was pointed out in Refs.\cite{VAPOR}, an analysis of quantum
evaporation from the condensate can yield valuable information
about the interatomic interaction. Therefore, adopting this 
analysis to the trapped gases is highly desirable. 
The very recent theoretical \cite{SUR} and experimental 
\cite{CORNELL} analyses of a Bose-Einstein condensate undergoing  
variations of the trapping potential address the long standing
question regarding the coherent versus dissipative behavior of
a many body system.  

The trap \cite{PETRICH} where the Bose-Einstein
 condensation of Rb atoms was 
first achieved \cite{ANDERSON}, utilizes
a rapidly rotating magnetic field (RMF). This field, if averaged
over the rotational period, creates an 
effective static oscillator
 potential (which is called the time orbiting potential (TOP) 
\cite{PETRICH}) and reduces the escape of atoms from the trap
due to spin-flip effects \cite{PETRICH}. 
Other traps \cite{BRADLEY,DAVIS} (see also \cite{SCI})
 do not rely on the RMF.  A main assumption made about the RMF
is that as long as the frequency of rotations $\omega$ is much larger than
the frequency $\omega_o$ of oscillations in the TOP, the trapped
atoms are not disturbed by the time variations of the instantaneous
potential.  
Accordingly, the  
results \cite{TRAP,BAYM} obtained for the Bose-Einstein condensation
in a static parabolic potential can be applied to this case as well. 

Generally speaking, the RMF
should transfer energy and angular momentum to the  condensate. 
Therefore, the RMF can be viewed 
as a possible tool for
studying the dynamical response of the condensate. In
this sense, addressing the problem of the exact description
of the quantum atomic states in the trap employing the RMF,
rather than
relying on the time averaging procedure \cite{PETRICH},
 appears to be
quite important.

   In this paper we study various aspects of the RMF: 
we find the exact single particle states in the trap \cite{PETRICH}
without relying on the time-averaging procedure;
it is shown that the RMF, if properly modulated, should excite
selectively some modes of the condensate;  
we derive the 
Ginzburg-Gross-Pitaevskii (GGP) equation taking
into account the effects of the quantum evaporation induced
by the RMF. As an application, the decay rate of 
the condensate
due to the RMF is calculated in the case of  steady rapid rotations of the
RMF.

The outline of the paper is as follows. In Section II 
we find the exact eigenenergies and eigenstates for a
single particle in the trap \cite{PETRICH}. This solution is obtained 
in the frame rotating together with the RMF. The
properties of the solution as a function of $\omega$ are analyzed.
 In  Sec.III we consider the limit of large numbers of particles in the 
condensate and analyze the nondissipative
interaction between the RMF and the normal modes of the condensate. 
In  Sec.IV the GGP equation with dissipation due to the RMF
is derived under certain approximations. 
The quasistatic solution for
the rate of the centrifugal evaporation of the condensate is derived in
the limit of high $\omega$.

  \section{Single Particle States in the rotating frame}
  
In our analysis of the behaviour of a single atom trapped by
the magnetic field $\bf{B}$ we follow the approximation that
the atomic spin orientation is parallel to $\bf{B}$
\cite{PETRICH}. Then
the effective potential energy of the atom seeking the low field is
essentially the Zeeman energy $U=|\mu_B \bf{B}|$, where $\mu_B $ stands
for the Bohr magneton (we ignore the nuclear magnetic moment). The magnetic
field of the trap \cite{PETRICH} consists of the static quadrupolar part
${\bf B}_q$ having axial symmetry with respect to the $z$-axis, and
the RMF ${\bf B}_b$ rotating in the $x,y$ plane.
 Representing these explicitly, one finds the components   
   
   \begin{equation} \begin{array}{l}
    B_{qx}=B'_qx,\,\, B_{qy}=B'_qy,\,\, B_{qz}= - 2B'_qz,   \\ \\ 
   B_{bx}=-B_b\cos (\omega t), \,\,
 B_{by}=-B_b\sin (\omega t),\,\, B_{bz}=0,
   \end{array} \end{equation}
      \noindent
   where $ B'_q,\,\, B_b$ stand for the constant gradient of the quadrupolar 
field, and the constant amplitude of the RMF, respectively.
Given Eq.(1), the potential energy
\begin{equation}
U=|\mu_B({\bf B}_q+{\bf B}_b)|
\end{equation}
\noindent
 depends on time. As suggested in Ref. \cite{PETRICH}, for high $\omega$
the time dependence can be effectively averaged over, which results in the
TOP-potential \cite{PETRICH}. In general, the time dependence 
in $U$ should result also in the nonadiabatic exchange of energy between
 the external field and the atoms in the trap. However, as will become
 clear from the following, for a time independent $\omega$ and 
noninteracting atoms, no such an exchange occurs between the RMF
and the atoms in the trap.
In fact, in the frame rotating together with the RMF, the
single-particle Hamiltonian becomes time independent insuring that the atom
once prepared in the pure state (in the rotating frame)
 will live forever in such a state. 
Going to the rotating frame implies
the coordinate transformation 

\begin{equation}\begin{array}{ll}
x''= & \cos(\theta (t))x +  \sin(\theta (t))y \\ \\ 
y'= & -\sin(\theta (t))x + \cos(\theta (t))y
\end{array}\end{equation}
\noindent
where $\theta (t)$ stands for the angle between
${\bf B}_b$ and the $x$-direction. 
In the case of steady rotations, $\theta (t)=\omega t$.
The transformation (3) results in (2) rewritten in the time 
independent form as

\begin{equation}
U=|\mu_BB'_q|\sqrt{(x''-x_o)^2+y'^2+4z^2},\,\,\, x_o={B_b \over B'_q}.
\end{equation}

Since the effective size of the atomic cloud is much less
then $x_o$ \cite{PETRICH}, one can expand (4) 
in terms of $1/x_o$. This gives for the
first two terms (linear and quadratic)

   \begin{equation}
U=\frac{\omega_{oy}^2y'^2}{2} + \frac{\omega_{oz}^2 z^2}{2} -x''  
 \end{equation}
   \noindent
where we have omitted the unimportant constant $|x_o|$; 
the notations $\omega_{oy}^2=1/|x_o|, \,\, \omega_{oz}^2=4/|x_o|$
are introduced, and the units of energy and length are employed as

\begin{equation}
\varepsilon_o= \frac{\hbar^2}{Ml^2_o},\,\, l_o=\frac{\hbar^{2/3}}
{(M|\mu_BB'_q|)^{1/3}},
\end{equation}
\noindent
respectively. In (6), $ M $ stands for the atomic mass. 
Note that in the rotating frame the stiffness of the potential along the
$x''$-coordinate is zero. This would imply that no states localized around
the origin exist. However, as we will see below, the finite kinetic energy
of the particle changes this conclusion for sufficiently large $\omega$.

  In the rotating frame the kinetic energy acquires the Coriolis term
$-\omega L$, where $L$ is the $z$ component of the angular momentum operator
and $\omega = \dot{\theta}$.
Consequently, taking into account (5), one finds the single particle 
Schr\"odinger
equation ($\hbar =1$) in the rotating frame 

\begin{equation}\begin{array}{l}
i\partial_t\psi=H_{\omega}\psi,\\   \\ 
H_{\omega}=-{1 \over 2} \Delta + i \omega (x''\partial_{y'} -
y'\partial_{x''}) +U
\end{array}\end{equation}

We consider first the case $\omega =const$.
Note that Eqs.(5), (7) represent a quadratic form which can be
diagonalized explicitly (see Appendix A). Prior to solving it
let us eliminate the linear term $-x''$ from (5). 
This can be accomplished by the transformation

\begin{equation}
\displaystyle \psi \Rightarrow {\rm exp}(-i{y' \over \omega})\psi (x',y',z),
\quad x'=x''-{1\over \omega^2}
\end{equation}
\noindent
which results in Eq.(7) being rewritten 
as

\begin{equation}\begin{array}{l}
i\partial_t\psi=H'_{\omega}\psi,\\   \\ 
\displaystyle H'_{\omega}=-{1 \over 2} \Delta + 
i \omega (x'\partial_{y'} -
y'\partial_{x'}) +    
 \frac{\omega_{oy}^2y'^2}{2} + 
\frac{\omega_{oz}^2 z^2}{2}' . 
\end{array}\end{equation}

For the case $|\omega| < \omega_{oy}$,
no discrete states localized near the origin $x'=y'=z'=0$ exist
in the trap. Accordingly, we will not analyze this case any more. For
$|\omega| > \omega_{oy}$ such states do exist. Their eigenenergies are
(see Appendix A)

\begin{equation}
\varepsilon_{mnl}=\omega_+m - \omega_-n+\omega_{oz}l, \quad 
\omega_{\pm}=\omega\sqrt{1+\eta^2/2 \pm \eta \sqrt{2+\eta^2/4}},
\quad \eta = \omega_{oy}/\omega ,
\end{equation}
\noindent 
where  $m, n, l$ are integer
nonnegative quantum numbers, and the energy of the state with
$m=n=l=0$ is set equal to zero.
The normalized eigenfunctions are (see Appendix A)

\begin{equation}
\psi_{mnl}(x',y',z)=\frac{(\omega_{oz}\omega_1 \omega_2)^{1/4}}
{\pi^{3/4}\sqrt{2^{m+n+l}m!n!l!}}e^{-\Xi_o}
\left [\frac{\partial^{m+n+l}}
{\partial t^m_1\partial t^n_2 \partial t^l_3}e^{\Xi}\right ]_{t_1=t_2=t_3=0},
\end{equation}
\noindent
where we have introduced

\begin{equation}\begin{array}{l}
\displaystyle \Xi_o=  \frac{\omega_1x'^2}{2}+
\frac{\omega_2y'^2}{2} + \frac{\omega_{oz}z^2}{2}- i\gamma_ox'y',  \\ \\
\displaystyle \Xi=\frac{\alpha^2-\alpha^{-2}}{2}(t^2_1-t^2_2) - t^2_3 -
2\nu t_1t_2 +2\sqrt{\omega_{oz}}t_3z +\\  \\ 
\displaystyle + \sqrt{2}[\alpha^{-1}
\sqrt{\omega_1}x' - i\nu\alpha \sqrt{\omega_2}y']t_1 +
\sqrt{2}[\nu\alpha
\sqrt{\omega_1}x' + i\alpha^{-1} \sqrt{\omega_2}y']t_2. 
\end{array}\end{equation}
\noindent
 In Eq.(12) the parameters are 

\begin{equation}\begin{array}{l}
\displaystyle \gamma_o=\omega \frac{\omega_2 - \omega_1}
{\omega_2 + \omega_1},\quad \omega_2 = \sqrt{1 - \eta^2}\omega_1,
\quad \eta = {\omega_{oy} \over \omega},
\quad \nu={\rm sign}( \omega ),  \\        \\  
\displaystyle \omega_1 = 
\sqrt{8 - 8\sqrt{1 - \eta^2} - 4\eta^2 - \eta^4}\eta^{-2}\omega,\quad 
 \alpha = \frac{(1 - {\eta^2 \over 4} - 
{\eta \over 4}\sqrt{8 + \eta^2})^{1/2}}{(1 + {\eta^2 \over 2}
 - {\eta \over 2}\sqrt{8 + \eta^2})^{1/4}( 1 - \eta^2)^{1/8}}.
 \end{array}\end{equation}

Note that the parameter
$\alpha$ as a function of $\omega$ (or $\eta$) has the
property $ \alpha (-\eta) \alpha (\eta) = 1$. In the limit
$\eta \to 0$ ($\omega \to \infty$) one obtains $ \alpha = 1 $,
and the eigenfunctions (11), (12) as well as 
the eigenenergies (10) become exactly those characterizing the
TOP \cite{PETRICH} as seen
from the rotating frame. In particular,
the spectrum acquires the form

\begin{equation}
\varepsilon'_{mnl}=\omega_o(m+n) - \omega (n-m) + \omega_{oz}l
\end{equation}
\noindent
where $\omega_o = \omega_{oy}/\sqrt{2}$ is the frequency
of the oscillations in the $x, y$-plane of the TOP
\cite{PETRICH}. 

For large 
$\omega$, Eq. (14) is an approximation
of the exact expression (10). The corrections due to the finiteness
of $\omega_o/\omega$ turn out to be of the order of 
$(\omega_o/\omega)^3$, so that
one can effectively ignore these even if $\omega$ is only a few times
larger than $\omega_o$. Consequently, for such $\omega$ we
will employ Eq.(14) instead of the exact form (10).

In the limit $ |\eta| \to 1$ from below, 
the solution (10)-(13) acquires features
characteristic of a quasi-1D system. Indeed, taking this limit in Eqs.
(10)-(13), one finds 

\begin{equation}
\omega_+= \omega_1 = \sqrt{3} \omega_{oy},\,\,\, \omega_-=
{1\over 3}\omega_2\approx
\sqrt{{2 \over 3}(1-|\eta|)}\omega_{oy}.
\end{equation}

This expression together with Eqs. (11), (12) imply 
 that the typical extension in the $y'$-direction diverges
as $(1 - |\eta |)^{-1/4} \to \infty$. Accordingly, the excitation
spectrum (10) becomes characterized by the soft
 mode  whose
energy $\omega_- $ goes to zero. This implies that in this region of 
$\omega$ the low energy dynamical response of the system of atoms 
should exhibit the 1D behavior. In this paper we will not focus on 
the properties of such a 1D atomic gas. 

Note that the spectrum (10) and its limiting form (14) have no
lower bound. For the model Hamiltonian whose potential is
the axially symmetric TOP \cite{PETRICH},
this is a pure consequence of the coordinate transformation (3) because
this Hamiltonian conserves angular momentum. 
Accordingly, no instability with respect to a spontaneous growth
of angular momentum $L=\hbar (n-m)$ (in physical units) in (14)
is expected to occur.
In contrast, the Hamiltonian (9) does not
conserve angular momentum. This implies that under certain conditions
such an instability could be realized.

Another interesting feature of the
solution (11), (12) is the common phase factor $exp(i\gamma_o x'y')$. 
Its magnitude is controlled by the asymmetry of the 
eigenfunctions in the $x,\, y$ plane (in addition to the squeezing in the
z-direction). Also, this factor implies a very specific
pattern for the velocity 
$ {\bf v} = \nabla {\rm Im}({\rm ln}(\psi_{00l})) $
at the levels with $ m=n=0 $. 
Employing (11), (12) for $m=n=0$ one finds

\begin{equation}
v_r= \gamma_o r \sin (2\theta),\,\,\, v_{\theta} = \gamma_o
r\cos(2\theta)
\end{equation}
\noindent
for the radial and polar components of ${\bf v}$, respectively 
(in the polar coordinates $ x'=r\cos(\theta),\,\,\,
y'=r\sin(\theta) $). This expression exhibits quadrupolar
symmetry. As long as particles are condensing into the state with
$m=n=l=0$ they will form the current pattern characterized by (16).
This pattern can be
thought of as two pairs of vortices of opposite vorticity
coupled together.

Above we have shown that 
no nonadiabatic
energy exchange occurs between the RMF and the ideal gas in the trap.
In the Sec.IV, we will show that the interaction between particles
changes this situation.

\section{condensate containing large numbers of particles
in the rotating frame} 

In this section we will analyze the case
of the condensate containing large numbers of particles in the 
presence of the RMF.
The condensate wave function
$\Phi$ obeys the  GGP equation \cite{LIF}. For
the single-particle Hamiltonian (7) in the rotating frame, this
equation is

\begin{equation}
i\partial_t \Phi = (H_{\omega} -\mu)\Phi + u_o|\Phi|^2\Phi   
\end{equation}
\noindent
where  
$u_o>0$ is the interaction
constant and $\mu$ stands for the chemical potential.
 Following the approach \cite{STRIN1}, we will derive
approximate hydrodynamical equations for the condensate in
the presence of the RMF.
We denote
\begin{equation}
\Phi=\sqrt{\rho}{\rm e}^{i\phi},\quad \int d{\bf x}\rho =N_c,
\end{equation}
\noindent
where $\rho,\> \phi$ and $ N_c$ are the density, the phase, and
the total number of particles in the condensate,
 respectively. Substituting (18) into (17), one arrives at 
the expressions

\begin{equation}\begin{array}{l}
\displaystyle \dot{\rho} - 
\omega (x''\partial_y'\rho - y'\partial_{x''}\rho) +
 \nabla (\rho \nabla \phi) =0,\\   \\ 
\displaystyle \dot{\phi} - 
\omega (x'\partial_y'\phi - y'\partial_{x''}\phi) =
 \frac{\Delta (\sqrt{\rho})}{2\sqrt{\rho}}
- {1\over 2}(\nabla \phi)^2 + \mu - U  - u_o\rho.
\end{array}\end{equation}
\noindent
in the rotating frame.
The main approximation made in the limit
of large $N_c$ is that the term proportional to $\Delta (\sqrt{\rho})$
in the second of equations (19)
can be neglected \cite{BAYM,STRIN1}. 

In the limit $\omega \to \infty$, one expects to obtain
a solution of (19) which is close to that characterizing
the TOP  
\cite{BAYM,STRIN1}. In order to see it, one should
separate a rotationally invariant part from the total potential
(5). Specifically, 

\begin{equation}\begin{array}{l}
U=U_{TOP} + \delta U,\\  \\ 
\displaystyle U_{TOP}={\omega^2_o \over 2}(x''^2+y'^2) + 
{\omega^2_{oz} \over 2}z^2,   
\quad \delta U= - {\omega^2_o \over 2}(x''^2-y'^2) - x'' .
\end{array}\end{equation}
\noindent
For the sake of convenience we will omit all primes
($x''\to x,\> y' \to y$) from the coordinates, implying that we are working
in the frame connected with the RMF unless
otherwise stated. Note that $U_{TOP}$ is the time averaged
potential (TOP) derived in Ref.\cite{PETRICH}. The term $\delta U$
describes the deviations of the instantaneous potential (5)
from $U_{TOP}$. It is not strictly obvious that $\delta U$ 
can be treated as a small correction to $U_{TOP}$.
 However, the exact results obtained above
for the single-particle Hamiltonian show that this is true in the 
limit of large $\omega$ at least. Below we will show that  
if $\omega >> \omega_o$, 
corrections to the solution (21) caused by $\delta U$
remain small for large $N_c$ as well. 

To the zeroth order with respect to $\delta U$, one obtains
from Eq.(19) the solution

\begin{equation}
\displaystyle \phi^{(o)}=0, \quad   
 \rho^{(o)} = {1\over u_o}(\mu - U_{TOP}), 
\end{equation}
\noindent
which
 is valid inside the droplet  whose
radius is determined by the condition $\rho^{(o)}=0$
\cite{BAYM,STRIN1}.  
We represent 
$\rho = \rho^{(o)} + \delta \rho$ \cite{STRIN1}, where
$\delta \rho$ is a small correction due to $\delta U$. Correspondingly,
we ignore the term $ \nabla (\delta \rho \nabla \phi)$ in the first equation
(19). 
Linearizing Eqs.(19) in $\delta \rho,\> \phi$ \cite{STRIN1}, one obtains

\begin{equation}\begin{array}{l}
\displaystyle \delta \dot{\rho} - 
\omega (x\partial_y\delta \rho - y\partial_x\delta \rho) +
 \nabla (\rho^{(o)} \nabla \phi) =0,\\  \\ 
\displaystyle \dot{\phi} -
 \omega (x\partial_y\phi - y\partial_x\phi) =
 - u_o\delta \rho - \delta U.
\end{array}\end{equation}

Note that these equations are a close analog to those 
obtained in \cite{STRIN1} for a trapping oscillator potential
which is spherically symmetric.
 The additional feature of 
Eqs.(22) is the term $\delta U$ which plays the role
of an external force. Later we will see that this term
under certain conditions can resonantly excite the condensate normal
modes with the angular momenta $L=1,\> 2$.

A particular solution corresponding to the symmetry
of the driving term $\delta U$ (20) can be taken in the form

\begin{equation}\begin{array}{l}
\delta \rho = \rho'_2(x^2-y^2) +2\rho''_2xy
+\rho'_1x+\rho''_1y,\\  \\ 
\phi=
\phi'_2(x^2-y^2) +2\phi''_2xy
+\phi'_1x+\phi''_1y,
\end{array}\end{equation}
\noindent
where $\rho'_l,\> \rho''_l,\> \phi'_l,\> \phi''_l, \> l=1, 2$ are
the time dependent amplitudes of the dipole ($l=1$), and
the quadrupole ($l=2$) harmonics. Substitution of (21) and (23)
into (22) yields

\begin{equation}\begin{array}{l}
\displaystyle \dot{\phi}_l + i l\omega \phi_l 
+ u_o\rho_l = ({\omega^2_o\over 2})^{l-1},\\  
\displaystyle \dot{\rho}_l + i l\omega \rho_l      
- l {\omega^2_o \over u_o}\phi_l =0.
\end{array}\end{equation}
\noindent
for the complex amplitudes

\begin{equation}
\displaystyle \rho_l = 
(\rho'_l + i\rho''_l){\rm e}^{il\omega t}, 
\quad \phi_l = 				
(\phi'_l + i\phi''_l){\rm e}^{il\omega t}.
\end{equation}
\noindent
Note that the dipole amplitudes $l=1$ describe
essentially the center of mass motion of 
the whole atomic cloud in the trap \cite{STRIN1}.

  If the RMF frequency $\omega$ does not change
in time, one obtains the steady solutions
($ \dot{\rho}=\dot{\phi}=0$) 

\begin{equation}
\displaystyle \phi'_1 =0,\quad \phi''_1= \frac{\omega}
{\omega^2_o - \omega^2},\quad
\rho'_1= \frac{\omega^2_o}			
{u_o(\omega^2_o - \omega^2)}, \quad \rho''_1=0,
\end{equation}
\noindent
and

\begin{equation}
\displaystyle \phi'_2 =0,\quad \phi''_2= -\frac{\omega^2_o}
{4\omega (1 - {\omega^2_o \over 2\omega^2 })},\quad
\rho'_2= \frac{\omega^2_o}{2u_o(2\omega^2 
 - \omega^2_o)},	
\quad \rho''_2=0.
\end{equation}
\noindent
This implies
that in the limit 
$\omega >> \omega_o$ the corrections due to the RMF
to the zeroth order solution (21) \cite{STRIN1,BAYM}
are small.

Note that Eqs.(27)
and (23) indicate that the phase factor ${\rm exp}(i\gamma_o
xy)$, discussed in the Sec.II for the ideal gas situation,
is not affected much by the interaction as long as $\omega
>> \omega_o$. Indeed, comparing Eqs.(11)-(13) with
Eqs.(23),(27), one finds that the parameter $\gamma_o$
in Eqs.(12), (13), (16) must be replaced by
$2\phi''_2 = \gamma_o + o((\omega_o/\omega)^3)$.
 When $\omega \to \omega_o$,
the solutions (26), (27) based on the condition
$\rho^{(o)}>>\delta \rho$ become no longer valid. 

We now consider the case when $\omega$
depends on time.
  For concreteness, we assume
that the frequency $\omega$ of the RMF is modulated
as

\begin{equation}
\omega =\overline{\omega} + \lambda \sin (\omega't) 
\end{equation}
\noindent
where $|\overline {\omega}| >> |\lambda|,$ and $\omega'$
are constants. Accordingly, one finds that the angle
between the RMF and the $x$-axis in the laboratory
frame is

\begin{equation}
\theta (t)= \overline{\omega}t + {\lambda \over \omega'}(1 - \cos (\omega't))
\end{equation}					

Employing the ansatz

\begin{equation}
\displaystyle \phi_l ={\rm e}^{-il\theta (t)}\tilde{\phi}_l,
\quad \displaystyle \rho_l ={\rm e}^{-il\theta (t)}\tilde{\rho}_l,
\end{equation}
\noindent
one obtains from Eqs.(24)

\begin{equation}
\displaystyle \tilde{\phi}_l=\frac{u_o}
{l\omega^2_o}\dot{\tilde{\rho}}_l,\quad
\displaystyle \ddot{\tilde{\rho}}_l + l\omega^2_o \tilde{\rho}_l = 
{\omega^{2l}_o \over u_o}{\rm e}^{il\theta (t)}.
\end{equation}			
\noindent
These equations indicate that
the resonance condition on $\omega'$ in (28) is
different for the dipole ($l=1$) and quadrupole ($l=2$)
harmonics. 
Indeed, given (29) and expanding the r.h.s. of the second
equation of  (31) 
in the small quantity $|\lambda /\omega'|<<1$, we get

\begin{equation}
\displaystyle {\rm e}^{il\theta (t)}\approx {\rm e}^{il\overline{\omega}t}
(1 + {il\lambda \over \omega}(1 - \cos \omega't)).		
\end{equation}
\noindent
Then, one obtains that
the resonance with the dipole harmonic occurs
when the modulating frequency $\omega'$ obeys
the condition

\begin{equation}
\omega'=\omega'_1=\overline{\omega} \pm \omega_o. 
\end{equation} 
\noindent
 Eqs.(31) yield the resonance condition 
for the quadrupolar harmonic

\begin{equation}
\omega'=\omega'_2= 2\overline{\omega} \pm \sqrt{2}\omega_o,
\end{equation} 			
\noindent 
where $\sqrt{2}\omega_o$ stands for the frequency
of the lowest quadrupolar harmonic of the trapped condensate
with large $N_c$ \cite{STRIN1}. Note that $\omega'_2 -
\omega'_1 \approx \overline{\omega} >> \omega_o$.

In the following, we will show
that the preceding analysis based on the GGP equation
does not take into account quantum processes of
the creation of pairs out of the condensate by the RMF.
These lead to forced evaporation of the condensate 
even for zero temperature and steady rotations
of the magnetic field. As a consequence, the
hydrodynamical equations (22) will acquire a dissipative
term.

\section{ Centrifugal Instabilty in the Many Body Approach}

In our previous analysis, we neglected
quantum fluctuations of the condensate. These fluctuations in
the conventional stable condensate
can be thought of as the virtual creation and absorption 
of pairs.
 In this regard we note that the spectrum (10)
 has no lower
bound, so that the condensate could be unstable with respect
to the real creation of pairs  even
though the RMF is steady. Correspondingly,
the GGP equation (17) can acquire a dissipative part. 

Consider first
the case $\omega \to \infty $. The 
term proportional to $ \omega$ in Eq.(14) 
is a direct consequence of the Galiliean transformation 
into the rotating frame. Indeed, the limit 
$\omega \to \infty$ in Eqs.(10)-(13) insures that the term 
$(n-m)$ is the projection of the angular momentum $L$ on the 
$z$-axis, so that the $\omega$-dependent part in (14) is exactly
the Coriolis contribution $ - \omega L$. This implies that no
instability should develop because the absence of the lower
bound for the spectrum is purely a frame of reference
effect. Nevertheless the condensate can be considered
as being 
potentially able to
gain high values of $L$. In this regard we can employ
the rotating frame reasoning \cite{LIF} (see also \cite{REV}, Ch.6)
 for the vortex creation in the
rotating vessel containing a superfluid. 
In the frame connected to the 
vessel rotating with 
the frequency $\omega$ around its axis, 
the vortex energy is $E_v - \omega L_v$ where $E_v$ and $ L_v$ stand for the
vortex energy in the laboratory frame 
and the vortex angular momentum, respectively. The vortex
can be created spontaneously if the Coriolis energy exceeds $E_v$. 
 However, this argument does not indicate what is the probability
for developing this centrifugal instability. In fact,
in the case of the perfectly symmetric vessel this probability
is essentially zero. To make the vortex creation real,
the vessel must have some irregularities on the walls breaking the
rotational symmetry so that the angular momentum of the vessel 
could be transferred to the vortex (or vortices). 

Returning to our case, we can see that 
in the case $\omega \to \infty $ the eigenfunctions (11), (12) of
the trap \cite{PETRICH} are approaching those of the effective time
averaged Hamiltonian (the TOP \cite{PETRICH})
 which is axially symmetric. 
 Therefore,
no centrifugal instability of the condensate is expected to occur in
this limit. In 
other words, no energy exchange between the RMF and the condensate happens
in the limit $\omega = \infty$.

For
finite $\omega$, the functions (11), (12) are not
eigenfunctions of the operator $L$.  
 Consequently, 
the difference $ n - m $ can no longer be interpreted as the 
eigenvalue of $L$. Accordingly the
effective vessel can be thought of as having a symmetry breaking
deformation, which in turn implies 
that the energy and the angular momentum can now be
given up to the pairs leaving the condensate
into the highly excited states whose energies
are $\hbar \omega >> \hbar \omega_o$
(in physical units).  In this regard one
should distinguish two cases: 1) $\hbar \omega >> \mu$ and
2) $\hbar \omega \leq \mu$. In the case 1) the pair escapes
into states lying far from those effectively involved in
the formation of the interacting condensate. Accordingly, 
the pair escape process can be treated as an incoherent step
in the condensate evaporation.
In contrast, in  case 2) the escape states with
the energies $\approx \hbar \omega$ are to be
renormalized strongly because of the presence of the
condensate. This implies that the multi-pair processes
become significant.
Correspondingly, the centrifugal
instability should be interpreted as a coherent
process of vortex formation. In this paper
we will not analyze this case.  

The process of the escape of pairs represents  
the nonresonant quantum evaporation of the
condensate induced by the RMF. 
We emphasize the crucial role of the interatomic
interaction for realization of this centrifugal evaporation. 
From the point of view of the rotating observer
this process can be described as follows: 
 two atoms in the condensate 
 interact with each other. As
a consequence, they
jump to a new pair of single-particle states characterized by
large quantum numbers, so that their total
energy is conserved. 
Correspondingly,
the rotating observer interprets this event as a nearly elastic
escape of the pair from the condensate. Note that if the
eigenfunctions (11) were eigenfunctions of angular momentum,
there would be a selection rule requiring that the angular momentum
of the interacting pair not change in the transition.
Correspondingly, referring to Eq.(14), one sees that
no instability would occur. In fact this is not the case 
for finite $\omega$ and instability could occur. 

To describe the centrifugal instability effect, we proceed
to derive a damping term in the GGP equation for the condensate
wave function $\Phi$. The many body Hamiltonian in the rotating
frame is

\begin{equation}
\displaystyle H=\int d{\bf x} [ \Psi^{\dagger}(H_{\omega} - \mu )\Psi +
{u_o \over 2}\Psi^{\dagger}\Psi^{\dagger}\Psi\Psi] ,
\end{equation}
\noindent
where primes are omitted from the coordinates and 
the Bose operators $\Psi^{\dagger},\,\,\, \Psi$
obey the usual Bose commutation rule. The Heisenberg equation
is

\begin{equation}
i\partial_t \Psi = (H_{\omega}- \mu)\Psi + u_o\Psi^{\dagger}\Psi\Psi.
\end{equation}

Taking into account the explicit form (7) for $H_{\omega}$,
 one finds from (36)
the current conservation condition 

\begin{equation}
\partial_t (\Psi^{\dagger}\Psi) + \nabla {\bf J} =0,
\end{equation}
\noindent
where the current operator
${\bf J}$ in the rotating frame is defined as

\begin{equation}\begin{array}{ll}
{\bf J}= & {1\over 2i}[ \Psi^{\dagger}(\nabla - {\bf A})\Psi],\\  \\ 
\displaystyle A_x=  & -i\omega y,\quad A_y=i\omega x, \quad A_z=0.
\end{array}\end{equation}

In the presence of the condensate, 
 the condensate wave function $\Phi = <\Psi>$. The noncondensate
part $\Psi'= \Psi - \Phi$. From Eq.(36), one 
finds \cite{BEL,FETTER,GRIF1}

\begin{equation}\begin{array}{c}
\displaystyle \partial_t \Phi = 
(H_{\omega}- \mu) \Phi + u_o|\Phi|^2\Phi +\\  \\
\displaystyle + u_o
[\Phi^*<\Psi' \Psi'> + 2 \Phi <\Psi'^{\dagger}\Psi'> + <\Psi'^{\dagger}
\Psi'\Psi'>],
\end{array}\end{equation}
\noindent
and

\begin{equation}\begin{array}{c}
\displaystyle i\partial_t \Psi' =  (H_{\omega} - \mu )\Psi'+ 
u_o[ \Phi^*(\Psi'\Psi' - <\Psi'\Psi'>)+ \\  \\
\displaystyle +2\Phi (\Psi'^{\dagger}\Psi' - <\Psi'^{\dagger}\Psi'>)  
 +2\Phi^*\Phi \Psi' + \Phi^2 \Psi'^{\dagger} + \Psi'^{\dagger}
\Psi' \Psi' - <\Psi'^{\dagger}\Psi'\Psi'>].
\end{array}\end{equation}

The condensate wave function $\Phi$ is normally viewed as an external classical
field in Eq.(40) for the noncondensate part. 
Employing
the Keldysh technique  \cite{KELD}, the system (39), (40) can
be expressed in terms of the joint dynamics of $\Phi $
and the population numbers of the excitations. In general, this procedure
is very complicated \cite{STOOF} (see also the generalized 
density functional approach \cite{GRIF,GRIF1}).
However, under certain conditions
it becomes possible to
 eliminate the averages from Eq.(39).
Specifically, we will make several assumptions and approximations:
a) 
 the population numbers of the excited states
are zero;
b)
 the pairs escaping from
the condensate due to the centrifugal effect escape from the trap
as well;
c) the terms leading to powers higher than third in $ \Phi$
and $\Phi^*$
in the effective GGP equation are omitted;
d) in Eq.(40) only the terms linear in $\Psi'$ and $\Psi'^{\dagger}$
are retained in accordance with the 
Bogolubov approximation \cite{LIF,FETTER}.

 The assumption a) excludes the normal component from
the analysis. The assumption b) insures that no normal component
is building up in the highly excited levels due to the centrifugal
escape of the pairs. Note that b) is reasonable
for high $\omega$ and in the presence of the radio-frequency scalpel
which provides the evaporative cooling (see, e.g., in \cite{COOL,BEC}).
 Given a) and b), we avoid the necessity to analyze the dynamics of the normal
component. 
Finally, from Eqs.(39), (40) under a)-d) 
we obtain the GGP equation with the dissipation
term included (see
Appendix B)

\begin{equation}
i\partial_t \Phi =( H_{\omega} - \mu)\Phi + u_o|\Phi|^2\Phi - i
u_o^2\Phi^*
\int dt'\int d{\bf x}'G^{(r)2}({\bf x}t,
{\bf x}'t')\Phi^2({\bf x}'t'),
\end{equation}
\noindent
where $G^{(r)}({\bf x}t, {\bf x}'t')$ is defined in (B7).
Note that if the total number of atoms in the trap were
conserved, it would not be possible to obtain the dissipation
term in (41) in closed form \cite{SUBIR}. Below it will be shown
explicitly that the last term in Eq.(41) would have been zero
if either the single particle Hamiltonian $H_{\omega}$ conserved
angular momentum or if the single particle excitation spectrum
were positively defined. 

  Multiplying (41) by $\Phi$ and adding the complex conjugate
of the resulting expression, one obtains 
the generalized current conservation condition

\begin{equation}\begin{array}{c}
\displaystyle \dot{ \rho} + {\rm div}{\bf J} = - S, \\   
\displaystyle S = u_o^2 \int dt'\int d{\bf x}'
G^{(r)2}({\bf x}t,{\bf x}'t')\rho ({\bf x}t)\rho 
({\bf x}'t'){\rm exp}[-2i(\phi ({\bf x}t) - \phi ({\bf x'}t'))] + c.c.,
\end{array}\end{equation} 
\noindent
where the representation
(18) is employed. This
equation corresponds to the first equation of Eqs.(22) modified to
include the
dissipation caused by the centrifugal evaporation.

Integration of Eq.(42) over the 
whole space yields 

\begin{equation}
\displaystyle \dot{N}_c = -2u_o^2 \int dt'd{\bf x}d{\bf x}'
{\rm Re}[G^{(r)2}({\bf x}t,{\bf x}'t')] \rho ({\bf x}t)\rho 
({\bf x}'t'){\rm exp}[-2i(\phi ({\bf x}t) - \phi ({\bf x'}t'))]
\end{equation} 
\noindent
where we have used the second relation in Eq.(18).
Note that in this equation the integral depends on both  
the density $\rho$ as well as the phase $\phi$ of the condensate. 
Accordingly, the coherence of the condensate could be tested by analyzing
the rate of the quantum vaporization induced by the RMF. Elsewhere, 
we will consider this possibility in greater detail.
Presently, let us calculate the quasistatic decay rate 
assuming that $N_c$ is not large so that
one can employ the ideal gas ansatz $\Phi = \sqrt{N_c}\psi_{ooo}$.
Note however that
the applicability of Eqs.(41)- (43) is not limited by the
requirement of small $N_c$ 
($a\sqrt{\omega_o}N_c\leq 1$ \cite{BAYM}, where $a$ is defined in (B5)). 
When the condition (B5) is satisfied,
case 1) holds, and
one can employ the variational approach \cite{BAYM}
for calculating $\Phi$. 

Making use of Eqs.(11), (12), we find 

\begin{equation}
\displaystyle \dot{N}_c(t)= - 2u^2_o\int dt'K(t-t')N_c(t)N_c(t')
\end{equation}
\noindent
where 

\begin{equation}
\displaystyle K(t-t') = \tilde{\theta}(t-t')
\sum_{1,2}|M_{12}|^2{\rm Re\> e}^{-i(\varepsilon_1 + 
\varepsilon_2)(t -t')},\quad
M_{12}= \int d{\bf x}\psi_1\psi_2
\psi^{*2}_{ooo}
\end{equation}
\noindent
and the summation is performed over the final states of the
escaping pair.

As will be seen below, in the limit $\omega \to \infty$
the escape rate is much smaller than the typical time scale
in the trap corresponding to $\omega_o$. Consequently, one can employ
the quasistatic approximation that the time dependence
of $N_c$ is slow. We set $N_c(t') = N_c(t)$ in (44) and
after the time-integration rewrite it as

\begin{equation}
\displaystyle \dot{N}_c= - \chi N_c^2 ,
\end{equation}
\noindent
where

\begin{equation} 
\displaystyle \chi =
2\pi u^2_o\sum_{[1],[2]}|M_{12}|^2\delta (E_{12}).
\end{equation}

In Eq.(47), the notation $E_{12}=\varepsilon_1 + \varepsilon_2$
for the escaping pair energy is introduced and $\delta(\xi)$ stands
for the $\delta$-function. This expression
accounts quantitatively for the centrifugal effect discussed above. One can
see that the condition $M_{12} \neq 0$ and $ E_{12}=0$, where explicitly

\begin{equation}
E_{12} = \omega_o(m_1 + m_2 + n_1 + n_2) + \omega_{oz} (l_1 + l_2)
- \omega (n_1 + n_2  - m_1 - m_2)=0,
\end{equation}
\noindent
can be satisfied simultaneously because the eigenfunctions (11), (12)
utilized
in (45) and (47) are not the
eigenfunctions of the angular momentum operator. In what follows,
we will show that in the limit of large $\omega$, the dominant 
contribution in (47) comes from the states with the quantum numbers

\begin{equation}
m_1\approx m_2 \approx l_1 \approx l_2 \approx n_1 \approx n_2 \approx
{\omega \over \omega_o} >> 1
\end{equation}
\noindent
corresponding to a pair leaving the condensate into the
states characterized by large quantum displacements. Returning to
the laboratory frame, this simply means that two atoms absorb the 
energy $2\hbar \omega$ (in physical units) 
from the RMF so that this energy is approximately equally
distributed between them. As a result, the pair is transferred
to highly excited states.
 The radio-frequency scalpel \cite{BEC,COOL} is assumed to 
eventually remove this pair insuring
the condition a) of zero population of the excited states. 

In order to
calculate $M_{12}$ explicitly, we employ the representation
(compare with \cite{FETTER}, Ch.15)

\begin{equation}
\displaystyle \psi_{m_1n_1l_1}({\bf x}_1)
\psi_{m_2n_2l_2}({\bf x}_2) = \sum_{pkq }
g^{m_1m_2}_pg^{n_1n_2}_kg^{l_1l_2}_q\psi_{pkq}({\bf R})
\psi_{p'k'q'}({\bf r})
\end{equation}
\noindent
where the notations
$p'= m-p,\> k'=n-k,\> q'=l-q$ and

\begin{equation}\begin{array}{l}
\displaystyle g^{ab}_{p}=\sum_{p'}(-1)^{p+p'}\frac{\sqrt{a!b!p!(a+b-p)!}}
{2^{{a+b \over 2}}p'!(a-p')!(p-p')!(b-p+p')!},\\    \\
{\bf R}=\frac{{\bf x}_1+{\bf x}_2}{\sqrt{2}},\quad 
{\bf r}=\frac{{\bf x}_1-{\bf x}_2}{\sqrt{2}},\quad 
m=m_1+m_2,\,\, n=n_1+n_2,\,\, l=l_1+l_2
\end{array}\end{equation}
\noindent
are employed. The summations in Eqs.(50), (51) run over all integer 
nonnegative numbers obeying the condition that all the numbers 
under the signs of the factorial  are nonnegative as well. The relations
(50), (51) were derived from the explicit representation (11),(12)
for the eigenfunctions.
Employing Eqs.(50), (51) in Eqs.(45),(47) we obtain for Eq.(47)

\begin{equation}
\displaystyle \chi  = {\pi u_o^2 \over 4}
\sum_{mnl}|\psi_{mnl}(0)|^2|\psi_{000}(0)|^2 \delta(
\omega_o(m+n)+\omega_{oz}l-\omega(n-m)).
\end{equation} 

Note that $\chi$ is exactly zero for the case $\alpha =1$ 
(or $\omega \to \infty$) in Eqs.(11)-(13). As 
mentioned above, no escape of the pairs occurs in the case when the RMF is
so rapidly rotating that the effective trapping (TOP) potential
becomes axially symmetric. In the limit of large but still finite
$\omega$ one finds from Eq.(13)

\begin{equation}
\alpha^2-\alpha^{-2}=-\left (\omega_o \over \omega \right)^3 +
o\left(
\omega_o \over \omega \right)^5
\end{equation}
\noindent
which implies that the first term only should
be kept in (53), and that the exponent ${\rm e}^{\Xi}$ in Eq.(11)
can be expanded in terms of the smallness of $\omega_o/\omega$.
 This expansion
represents the eigenfunction (11) in terms of the harmonics of the
angular momentum operator. Keeping the first term only, one finds
that each eigenfunction can be effectively characterized by
three terms: $i$) the harmonic of the angular momentum operator with
the angular momentum $(n-m)$; $ii$) two functions with the momenta
$ (n-m) \pm 2$ whose weight is proportional to $(\omega_o/\omega)^3 << 1$. 
This implies that in Eq.(47) it is enough to consider
the contribution due to the lowest term. Physically
this term corresponds to an absorption of the energy $2\hbar \omega$
and the angular momentum $2$ by a pair of atoms escaping from the
condensate. Finally one finds

\begin{equation}
\displaystyle \chi ={1\over 4} a^2\omega_o^2\omega_{oz}
\left ({\omega_o \over \omega}\right )^6\sum_{ml}
\frac{(2l)!(m+1)(m+2)}{2^{2l}(l!)^2}\delta (2\omega_om+
2\omega_{oz}l - 2\omega).
\end{equation}
\noindent
 A simple analysis shows that most of the contribution to the sum (54)
comes from the region of high $m,\,l$
(see (49)). Accordingly, we replace the summation in (54) by integration.
 Finally, in the chosen limit and chosen units (6) we find

\begin{equation}
\displaystyle \chi =\epsilon a^2\omega_o^2
\left ({\omega_o \over \omega}\right )^{7/2},\quad
\epsilon = \frac{2^{7/4}}{15 \sqrt{\pi }} \approx 0.13. 
\end{equation}

This expression indicates that the centrifugal escape rate is extremely
sensitive to the RMF frequency $\omega$. For the parameters
employed experimentally in Ref.\cite{ANDERSON}, the estimate
of (55) gives a very small number (the corresponding
lifetime is about $10^6s$ for $N_c=1000$),
implying that the centrifugal
vaporization can be effectively ignored as a cause for
the condensate escape from the trap. However, with decrease of the ratio
$\omega /\omega_o$, the vaporization rate increases strongly. In the
case $\omega \to \sqrt{2}\omega_o $ from above, the approximation (53) we
employed is no longer valid. Accordingly, the exact expression
(11)-(13) for the eigenfunctions should be utilized in (52). This means
that the escaping pairs acquire  higher (even) angular momenta. As
a result, the lifetime of the condensate can become very short.

 Eqs.(39) and (40) can be analyzed for the case of a nonsteady
RMF. Especially interesting appears to be the 
case when the RMF excites resonantly the quadrupolar
harmonic of the condensate ( see the condition (34)).
Generally it is natural to expect
that this resonance would result in the increase of
the vaporization rate as a function of the modulating frequency
$\omega'$. In the future,
we will consider this case in greater detail.

\section{Conclusion}

The atomic trap \cite{PETRICH,ANDERSON} bears features 
absent in the static traps \cite{BRADLEY,DAVIS,SCI}. These features
can be accounted for in the frame rotating together with the RMF. For large
frequencies of rotation of the RMF, the exact eigenenergies and 
eigenfunctions of the trap \cite{PETRICH,ANDERSON}
 approach those characterizing the time
averaged potential TOP \cite{PETRICH}
 having axial symmetry with respect
to the axis of rotation. For frequencies close to the threshold
below which the trapping is impossible, the eigenstates loose
their axial symmetry and become elongated in the direction
perpendicular  to the RMF (in its frame of reference). Very close to
the threshold a gas of trapped atoms acquires properties of an essentially
1D system.

Due to the asymmetry introduced by the RMF, the atom-atom interaction
results in the induced evaporation of the Bose-Einstein condensate.
The time
scale for this evaporation is very sensitive to the RMF frequency of
rotation. For high frequencies, the lifetime of the condensate
increases as a large power of $\omega$. Close to the trapping threshold
the lifetime shortens considerably, implying that the 1D gas formed
in the trap \cite{ANDERSON,PETRICH}
 in this situation is a strongly interacting system. 

The RMF can be utilized as a driving force selectively exciting
the condensate normal modes. In the limit of large numbers of atoms,
 when Stringari's hydrodynamical approximation is valid, two modes
can be excited by the RMF whose frequency of rotation is 
appropriately
modulated. The first is a dipole mode accounting for the center
of mass motion. The second mode which can be excited by the RMF
is the lowest quadrupolar harmonic. The resonance conditions for the RMF
modulation period depend on the averaged RMF frequency,
in addition to the eigenfrequencies of the harmonics.
 The effect of the quantum evaporation induced by the RMF
opens up a channel for dissipation of the condensate normal modes.

\acknowledgments

This research was supported by grants from The City University
of New York PSC-CUNY Research Award Program.

\appendix
\section{Solving the eigenproblem in the rotating frame}

The Hamiltonian (9)  for the function
$\psi'(x',y',z)={\rm exp}(i\omega x''y') \psi$ can be written as

\begin{equation}\begin{array}{l}
\displaystyle H'_{\omega} = H_{xy} + H_z,\quad
 H_z= -{1\over 2}\partial_z^2 + 
{\omega^2_{oz} \over 2}z^2,\\  \\ 
\displaystyle H_{xy}=-{1\over 2}(\partial^2_x + \partial^2_y) +
\frac{\omega^2_{oy} - \omega^2}{2}y^2 + {3 \over 2}\omega^2 x^2
+2i\omega x\partial_y 
\end{array}\end{equation}
\noindent
where  
the unimportant constant $1/(2\omega^2)$ 
and the primes from the coordinates are omitted. The
eigenfunctions $\varphi_l (z)$ of $H_z$ 
are the well known oscillator states. These are
represented in (11), (12) by the generating function of the
Hermite polynomials (see the auxiliary variable $ t_3 $ in (11), (12))
 so that the $\psi'(x,y,z) = \varphi_l (z) \psi'_{mn}(x,y)$. Performing
the Fourier transform

\begin{equation}
\displaystyle \tilde{\psi}(x,p)=\int d\tilde{y}e^{-ip\tilde{y}}
\psi'(x,\tilde{y}), \quad \tilde{y}= 
\sqrt{|\omega^2 - \omega^2_{oy}|}\,y,
\end{equation}
\noindent
one finds for (A1) 

\begin{equation}
\displaystyle H_{xy}= - {1\over 2}\partial^2_x + 
{s\over 2}\partial^2_p + \frac{\omega^2 - \omega^2_{oy}}{2}
p^2 + {3\over 2}\omega^2x^2 - 2\omega \sqrt{|\omega^2 - 
\omega^2_{oy}|}\,  xp
\end{equation}
\noindent
where $ s = {\rm sign}(\omega^2- \omega^2_{oy})$.

In the case $|\omega| < \omega_{oy}$ the Hamiltonian (A3) can be 
diagonalized by implementation of a real rotation in the $(x, p)$-plane.
However, no discrete states exist in this case because the effective
potential of (A3) turns out to have a saddle like shape, with
the kinetic part being positively defined. In the 
opposite limit ( $ s=1$) the discrete states do exist. The
diagonalization can be achieved by means of the Lorentz transformation

\begin{equation}\begin{array}{ll}
x=& \cosh (\vartheta) \xi + \sinh (\vartheta) p'\\  \\
p=& \sinh (\vartheta) \xi + \cosh (\vartheta) p'
\end{array}\end{equation}
\noindent
leaving the kinetic part $ - { 1 \over 2 } (\partial^2_x-\partial^2_p)$
 invariant. In terms of the new variables ($\xi, p'$), (A3) acquires
the form

\begin{equation}
\displaystyle H_{xy} = - {1\over 2}\partial^2_{\xi} 
+ {\omega^2_+ \over 2}
 \xi^2 - [ - {1\over 2}\partial^2_{p'} +
{\omega^2_-\over 2}p'^2] 
\end{equation}
\noindent
where $\omega_{\pm} $ are given in Eq.(10), and the angle $\vartheta$
satisfies the equation

\begin{equation}
\displaystyle \tanh (2\vartheta)=\frac{4\nu \sqrt{1 - \eta^2}}
{4 - \eta^2}, 
\end{equation}
\noindent
with $\nu,\> \eta$ defined in Eq.(13). The resulting spectrum of the total Hamiltonian
 is given by Eq.(10). The eigenfunctions of (A5),
expressed in terms of the $\xi, p'$ variables, can be converted into
the $ x, p$ coordinates by means of the relations (A4). Finally, performing
the inverse of the Fourier as well as the scaling transforms (A2), one finds
the normalized eigenfunctions (11) - (13).

\section{derivation of the GGP equation with the dissipation due to the RMF}

Under a)-d), Eqs.(39) and (40) simplify considerably. We need to
find the lowest order term which contributes to the imaginary
part of Eq.(39). 
In Eq.(39), the
second term in the brackets does not contribute to the imaginary
part, so we omit it. The last term in the brackets of 
Eq.(39) produces the imaginary part. However it
can be shown that it is proportional 
to the population numbers of the excited states. Consequently, 
we omit this term also and rewrite (39) as

\begin{equation}
\displaystyle i\partial_t \Phi =( H_{\omega} - \mu)\Phi + u_o|\Phi|^2\Phi + u_o
\Phi^*G({\bf x}t,{\bf x}t)
\end{equation}
\noindent
where the equal-time anomalous Green's function \cite{BEL,FETTER} is 
defined as $G({\bf x}t,
{\bf x}'t)=<\Psi'({\bf x}t)\Psi'({\bf x}'t)>$ ( the overall
factor $i$ is omitted
here and below in the Green's functions definitions \cite{BEL,FETTER}).
 Note that
because the interaction potential in (35) is chosen in
the $\delta ({\bf x})$ form, Eq.(B1) contains $G$ with ${\bf x}={\bf x}'$.
The equation for $G$ can be obtained from Eq.(40). 
In order to accomplish this, we
will employ the Bogolubov approximation d).
Accordingly, multiplying (40) by
$\Psi'$ and taking the average, one finds

\begin{equation}\begin{array}{c}
\displaystyle i\partial_t G({\bf x}t, {\bf x}'t)=(H_{\omega} - \mu)_{\bf x}
G({\bf x}t, {\bf x}'t) + (H_{\omega} - \mu)_{{\bf x}'}
G({\bf x}t, {\bf x}'t) +\\   \\ 
\displaystyle + u_o[ 2(\Phi^*({\bf x}t)\Phi({\bf x}t) + 
 \Phi^*({\bf x}'t)\Phi({\bf x}'t)) G({\bf x}t,{\bf x}'t) +\\   \\ 
\displaystyle + \Phi^2({\bf x}'t) G^{-+}({\bf x}t,{\bf x}'t) +
 \Phi^2({\bf x}t)G^{+-}({\bf x}t,{\bf x}'t)],
\end{array}\end{equation}
\noindent
where the normal Green's functions for coinciding times are

\begin{equation}\begin{array}{l}
G^{-+}({\bf x}t, {\bf x}'t) =
<\Psi'({\bf x}t)\Psi'^{\dagger}({\bf x}'t)>, \\   \\ 
G^{+-}({\bf x}t, {\bf x}'t) =<\Psi'^{\dagger}({\bf x}t)\Psi'({\bf x}'t)>.
\end{array}\end{equation}
\noindent
 In Eq.(B2) the notation $(...)_{\bf x}$ means that
 the single particle Hamiltonian $ H_{\omega}$ acts on
the coordinate ${\bf x}$. 
These equations should be supplemented by ones for the normal
Green's functions. However, as long as one is only interested in deriving
the imaginary contribution to (B1) to lowest order
with respect to $u_o$, significant simplification can be achieved. 
Moreover, for $\omega >> \omega_o$, 
 only the high energy part of
the spectrum of the normal excitations contributes to the
imaginary part of $G$ in Eq.(B1). Correspondingly, one
can neglect the effect of the condensate on this part
of the spectrum (see the case 1) discussed above).
 The condition when this assumption is valid
can be formulated in terms of the smallness of the first term
in the square brackets of Eq.(B2) if compared with $\omega G$.
In other words,

\begin{equation}
\hbar|\omega| >> u_o|\Phi|^2
\end{equation}
\noindent
in physical units. Employing the variational approach \cite{BAYM},
one can estimate $|\Phi|^2$ and obtain from (B4) 

\begin{equation}
\displaystyle a\sqrt{\omega_o} N_c\leq 0.034\left (
{\omega \over \omega_o}\right )^{5/2}
\end{equation}
where we have employed the representation $u_o=4\pi a$ for
the interaction constant $u_o$ in terms of the scattering length
$a$ in the units (6). Actually, for the parameters
of the trap \cite{ANDERSON} the estimate for $N_c$ gives
$N_c \leq 10^6$.  
If this condition holds, the anomalous Green's function $G$ can
be found by iteration with respect to $u_o$, with the
zeroth order approximation being zero. Correspondingly, the 
normal functions (B3)
$G^{+-}$ and $G^{-+}$ should be taken in the zeroth order as
$G^{+-} = 0$ and   
$G^{-+}= \delta ({\bf x} - {\bf x}')$ (see conditions a) and b)). 
Assuming that (B4) (or (B5)) is valid, we finally find

\begin{equation}
G({\bf x}t, {\bf x}t)= - iu_o\int dt'\int d{\bf x}'G^{(r)2}({\bf x}t,
{\bf x}'t')\Phi^2({\bf x}'t')
\end{equation}
\noindent
where the retarded Green's function \cite{FETTER}
\begin{equation}
G^{(r)}({\bf x}t,
{\bf x}'t')= \tilde{\theta} (t-t') \sum_{1} 
{\rm e}^{-i\varepsilon_1(t-t')}\psi_1
({\bf x})\psi^*_1({\bf x}')
\end{equation}
\noindent
is expressed explicitly in terms of the single-particle
eigenfunctions (11), (12) and the eigenvalues
(10), with the summation performed over
all the single-particle quantum numbers indicated as $1$.
In (B7), $\tilde{\theta}(\tau)$ denotes the step function. 
Finally, substitution of Eqs.(B6) and (B7) into Eq.(B1)
yields Eq.(41).


\begin{references}
  \bibitem{ANDERSON}
 M.H. Anderson, J.R. Ensher, M.R. Matthews, C.E. Wieman, 
and E.A. Cornell, Science {\bf 269}, 198 (1995).
  \bibitem{BRADLEY}
 C.C. Bradley, C.A. Sackett, J.J. Tollett, and R.G. Hulet,
 Phys. Rev. Lett.{\bf 75}, 	1687 (1995).
   \bibitem{DAVIS}
 K.B. Davis, M.-O. Mewes, M.R. Andrews, N.J. van Druten, 
D.S. Durfee, 
D.M. Kurn, and W. Ketterle, Phys. Rev. Lett. {\bf 75}, 3969 (1995).
\bibitem{BEC}
{\it Bose-Einsten Condensation}, ed. by A. Griffin, D.W. Snoke,
S. Stringari (Cambrige University Press, Cambridge, 1995).
\bibitem{SUBIR}
 B.I. Halperin, P.C. Hohenberg, E.D. Siggia, Phys.Rev.{\bf B 13}, 1299 (1976);
K. Damle, S.N. Majumdar and S. Sachdev,  cond-mat/9511058.
\bibitem{KAGAN}
Yu. Kagan, in \cite{BEC}, p. 202.
\bibitem{STOOF}
H.T.C. Stoof, in \cite{BEC}, p. 226.
  \bibitem{STRIN1}
S. Stringari,  cond-mat/9603126.
   \bibitem{SHLAP}
B.V. Svistunov and G.V. Shlyapnikov, Sov.Phys.JETP, {\bf 71}, 71 (1990);
H.D. Politzer, Phys.Rev.A {\bf 43}, 6444 (1991);
J. Javanainen, Phys. Rev. Lett., {\bf 75}, 1927 (1995).
\bibitem{REV}
D.R. Tilley and J. Tilley, {\it Superfluidity and
 Superconductivity} ( Adam Hilger,
Bristol, 1990).
\bibitem{STRIN2}
S. Stringari, Phys. Rev. Lett.{\bf 76}, 1405 (1996).
\bibitem{VAPOR}
P.A. Mulherman and J.C. Inkson, Phys. Rev. B {\bf 46}, 5454 (1992);
F. Dalfovo, A. Fracchetti, A. Lastri, L. Pitaevskii, and S. Stringari,
Phys. Rev. Lett. {\bf 75}, 2510 (1995).
\bibitem{SUR}
Yu. Kagan, E.L. Surkov, and G.V. Shlyapnikov, 
 atom-ph/9606001.
\bibitem{CORNELL}
M. Matthews, D. Jin, J. Ensher, C. Wieman, E. Cornell, Summaries 
(QPD9-2) 
of QELS '96, Anaheim, California, June 2-7, 1996.
   \bibitem{PETRICH}
 W. Petrich, M.H. Anderson, J.R. Ensher, E.A. Cornell,
 Phys. Rev. Lett.{\bf 74}, 3352 (1995).
\bibitem{SCI}
 G. Taub, Science {\bf 272}, 1587 (1996).
   \bibitem{TRAP} 
V.V. Goldman, I. Silvera, A.J. Leggett, Phys. Rev. B {\bf 24}, 2870 (1981);
V. Bagnato, D.E. Pritchard, and D. Kleppner, Phys. Rev. A {\bf 35}, 4354 (1987).
   \bibitem{BAYM}
G. Baym and C.J. Pethick, Phys. Rev. Lett. 76, 6 (1996).
\bibitem{LIF}
E.M. Lifshitz and L.P. Pitaevskii, {\it Statistical Physics,
Part 2} (Pergamon Press, Oxford, 1980).
\bibitem{BEL}
S.T. Beliaev, Sov.Phys. JETP {\bf 7}, 289 (1958).
\bibitem{FETTER}
A.L. Fetter and J.D. Walecka, {\it Quantum Theory of Many-Particle Systems}
(McGraw-Hill, New York, 1971).
\bibitem{GRIF1}
A. Griffin,  cond-mat/9602036.
   \bibitem{KELD}
L.V. Keldysh, Sov. Phys. JETP {\bf 20}, 1018 (1965);
L.V. Keldysh in \cite{BEC}, p. 246.
\bibitem{GRIF}
A. Griffin, Can. J. Phys. {\bf 73}, 755 (1995). 
\bibitem{COOL}
K.B. Davis, M.-O. Mewes, M.A. Joffe, M.R. Andrews, W. Ketterle,
Phys. Rev. Lett. {\bf 74}, 5202 (1995).
\end{references}
   \end{document}